\documentclass[12pt]{article}

\usepackage{cite,times,fullpage,ifthen,graphicx}
\graphicspath{{figure/}}
\usepackage[font=sf, labelfont={sf,bf}]{caption}

\def\aj{Astron. J.}%
\def\apj{Astrophys. J.}%
\def\aap{Astron. Astrophys.}%
\def\mnras{Mon. Not. R. Astron. Soc.}%
\def\pasp{PASP}%
\def\nat{Nature}%
\makeatletter
\renewcommand\@biblabel[1]{#1.}

\def\@cite#1#2{$^{\mbox{\scriptsize #1\if@tempswa , #2\fi}}$}

\newcommand{\spacing}[1]{\renewcommand{\baselinestretch}{#1}\large\normalsize}

\def\@maketitle{%
  \newpage\spacing{1}\setlength{\parskip}{12pt}%
    {\Large\bfseries\noindent\sloppy \textsf{\@title} \par}%
    {\noindent\sloppy \@author}%
}

\newenvironment{affiliations}{%
    \setcounter{enumi}{1}%
    \setlength{\parindent}{0in}%
    \slshape\sloppy%
    \begin{list}{\upshape$^{\arabic{enumi}}$}{%
        \usecounter{enumi}%
        \setlength{\leftmargin}{0in}%
        \setlength{\topsep}{0in}%
        \setlength{\labelsep}{0in}%
        \setlength{\labelwidth}{0in}%
        \setlength{\listparindent}{0in}%
        \setlength{\itemsep}{0ex}%
        \setlength{\parsep}{0in}%
        }
    }{\end{list}\par\vspace{12pt}}

\renewenvironment{abstract}{%
    \setlength{\parindent}{0in}%
    \setlength{\parskip}{0in}%
    \bfseries%
    }{\par\vspace{-6pt}}

\renewcommand{\section}{\@startsection {section}{1}{0pt}%
    {-6pt}{1pt}%
    {\bfseries}%
    }
\renewcommand{\subsection}{\@startsection {subsection}{2}{0pt}%
    {-0pt}{-0.5em}%
    {\bfseries}*%
    }

\newenvironment{addendum}{%
    \setlength{\parindent}{0in}%
    \small%
    \begin{list}{Acknowledgements}{%
        \setlength{\leftmargin}{0in}%
        \setlength{\listparindent}{0in}%
        \setlength{\labelsep}{0em}%
        \setlength{\labelwidth}{0in}%
        \setlength{\itemsep}{12pt}%
        }
    }
    {\end{list}\normalsize}

\newcommand{\NAT@ignore}[2][]{}

\makeatother

\newcommand\fd{\hbox{$.\!\!^{\reset@font\romn d}$}}
\newcommand\fh{\hbox{$.\!\!^{\reset@font\romn h}$}}
\newcommand\fm{\hbox{$.\!\!^{\reset@font\romn m}$}}
\newcommand\fs{\hbox{$.\!\!^{\reset@font\romn s}$}}

\newcommand\farcs{\hbox{$.\!\!^{\prime\prime}$}}
\newcommand\fp{\hbox{$.\!\!^{\reset@font\reset@font\scriptscriptstyle\romn p}$}}
\renewcommand{\deg}{\hbox{$^\circ$}}
\newcommand{\arcmin}{\hbox{$^\prime$}}

\title{A filament of dark matter between two clusters of galaxies}

\author{J\"org P. Dietrich$^{1}$, Norbert Werner$^{2}$, Douglas
  Clowe$^3$, Alexis Finoguenov$^4$, Tom Kitching$^5$, Lance Miller$^6$
  \& Aurora 
  Simionescu$^2$}

\spacing{1}

\newcounter{firstbib}

\begin{document}

\maketitle

\begin{affiliations}
\item Physics Dept. and Michigan Center for Theoretical Physics,
  University of Michigan, 450 Church Street, Ann Arbor, MI 48109-1040,
  USA
 \item Kavli Institute for Particle Astrophysics and Cosmology,
   Stanford University, 382 Via Pueblo Mall, Stanford, CA 94305-4060,
   USA
\item Dept.of Physics \& Astronomy, Ohio University, Clippinger Lab
  251B, Athens, OH 45701, USA
\item Max-Planck-Institut f\"ur extraterrestrische Physik,
  Giessenbachstra{\ss}e, 85748 Garching b. M\"unchen, Germany
\item Institute for Astronomy, The University of Edinburgh, Royal
  Observatory, Blackford Hill, Edinburgh EH9 3HJ, U.K.
\item Department of Physics, University of Oxford, The Denys Wilkinson
  Building, Keble Road, Oxford OX1 3RH, U.K.
\end{affiliations}

\spacing{1}

\begin{abstract}
  It is a firm prediction of the concordance Cold Dark Matter (CDM)
  cosmological model that galaxy clusters live at the intersection of
  large-scale structure filaments\cite{1996Nature..380..603B}.
  The thread-like structure of this ``cosmic web'' has been traced by
  galaxy redshift surveys for
  decades\cite{1978MNRAS.185..357J,1989Sci...246..897G}. More recently
  the Warm-Hot Intergalactic Medium (WHIM) residing in low redshift
  filaments has been observed in emission\cite{2008A&A...482L..29W}
  and absorption\cite{2009ApJ...695.1351B,2010ApJ...714.1715F}.
  However, a reliable direct detection of the underlying Dark Matter
  skeleton, which should contain more than half of all
  matter\cite{2010MNRAS.408.2163A}, remained elusive, as earlier
  candidates for such
  detections\cite{1998astro-ph/9809268K,2002ApJ...568..141G,
    2005A&A...440..453D} were either
  falsified\cite{2004A&A...422..407G,2008MNRAS.385.1431H} or suffered
  from low signal-to-noise
  ratios\cite{1998astro-ph/9809268K,2005A&A...440..453D} and
  unphysical misalignements of dark and luminous
  matter\cite{2002ApJ...568..141G,2005A&A...440..453D}.
  Here we report the detection of a dark matter filament connecting
  the two main components of the Abell 222/223 supercluster system
  from its weak gravitational lensing signal, both in a non-parametric
  mass reconstruction and in parametric model fits. This filament is
  coincident with an overdensity of
  galaxies\cite{2002A&A...394..395D,2005A&A...440..453D} and diffuse,
  soft X-ray emission\cite{2008A&A...482L..29W} and contributes mass
  comparable to that of an additional galaxy cluster to the total mass
  of the supercluster. Combined with X-ray
  observations\cite{2008A&A...482L..29W}, we place an upper limit of
  $\mathbf{0.09}$ on the hot gas fraction, the mass of X-ray emitting
  gas divided by the total mass, in the filament.
\end{abstract}

\begin{figure}
  \includegraphics[width=\textwidth]{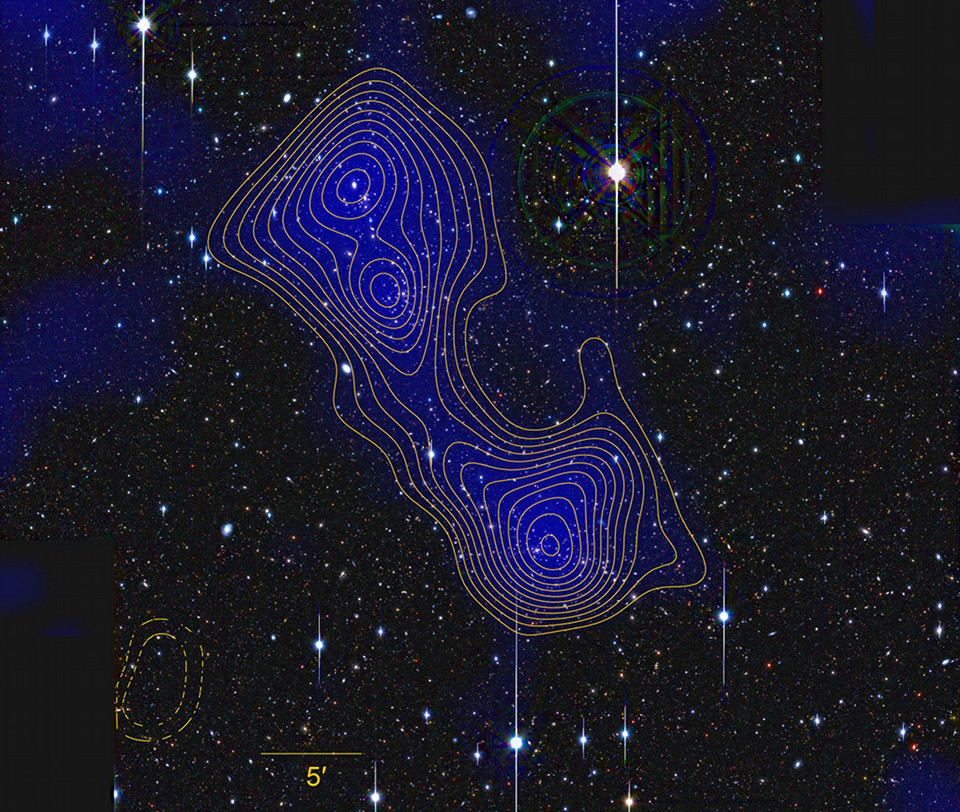}
  \caption{\textbf{Mass reconstruction of A~222/223.} The background
    image is a three colour-composite SuprimeCam image based on
    observations with the 8\,m Subaru telescope during the nights of
    Oct. 15 (A~222) and 20 (A~223), 2001 in V-, R$_\mathrm{c}$- and
    i$^\prime$-bands. We obtained the data from the SMOKA science
    archive (\texttt{http://smoka.nao.ac.jp/}). The FWHM of the
    stellar point-spread function varies between $0\farcs57$ and
    $0\farcs70$ in our final co-added images. Overlayed are the
    reconstructed surface mass density (blue) above $\kappa = 0.0077$,
    corresponding to $\Sigma = 2.36 \times
    10^{13}\,M_\odot$\,Mpc$^{-2}$, and significance contours above the
    mean of the field edge, rising in steps of $0.5\sigma$ and
    starting from $2.5\sigma$. Dashed contours mark underdense regions
    at the same significance levels. Supplementary Figure~1 shows the
    corresponding B-mode map. The reconstruction is based on 40,341
    galaxies whose colours are not consistent with early type galaxies
    at the cluster redshift.  The shear field was smoothed with a
    $2\arcmin$ Gaussian. The significance was assessed from the
    variance of 800 mass maps created from catalogues with randomised
    background galaxy orientation. We measured the shapes of these
    galaxies primarily in the R$_\mathrm{c}$-band, supplementing the
    galaxy shape catalogue with measurements from the other two bands
    for galaxies for which no shapes could be measured in the
    R$_\mathrm{c}$-band, to estimate the gravitational
    shear\cite{2007MNRAS.382..315M,2008MNRAS.390..149K}. A~222 is
    detected at $\sim 8.0\sigma$ in the south, A~223 is the
    double-peaked structure in the north seen at $\sim 7\sigma$. }
\label{fig:mass-reconstruction}
\end{figure}

Abell~222 and Abell~223, the latter a double galaxy cluster in itself,
form a
\makeatletter \def\@cite#1#2{#1\if@tempswa , #2\fi} \makeatother
supercluster system of three galaxy clusters at a redshift of $z \sim
0.21$ [ref.~\cite{2002A&A...394..395D}], separated on the sky by $\sim
14\arcmin$. Gravitational lensing
\makeatletter \def\@cite#1#2{$^{\mbox{\scriptsize #1\if@tempswa ,
      #2\fi}}$} \makeatother 
distorts the images of faint background galaxies as their light passes
massive foreground structures. The foreground mass and its
distribution can be deduced from measuring the shear field imprinted
on the shapes of the background galaxies. Additional information on
this process is given in the supplementary information. The mass
reconstruction in Figure~\ref{fig:mass-reconstruction} shows a mass
bridge connecting A~222 and the southern component of A~223 (A~223-S)
at the $4.1\sigma$ significance level. This mass reconstruction does
not assume any model or physical prior on the mass distribution.

To show that the mass bridge extending between A~222 and A~223 is not
caused by the overlap of the cluster halos but in fact due to
additional mass, we also fit parametric models to the three clusters
plus a filament component. The clusters were modelled as elliptical
Navarro-Frenk-White (NFW) profiles\cite{1997ApJ...490..493N} with a
fixed mass-concentration relation\cite{2004A&A...416..853D}. We used a
simple model for the filament, with a flat ridge line connecting the
clusters, exponential cut-offs at the filament end points in the
clusters, and a King profile\cite{1966AJ.....71...64K} describing the
radial density distribution, as suggested by previous
studies\cite{2005MNRAS.359..272C,2010MNRAS.401.2257M}. We show in the
supplementary information that the exact ellipticity has little impact
on the significance of the filament.

\begin{figure}
  \includegraphics[width=\textwidth]{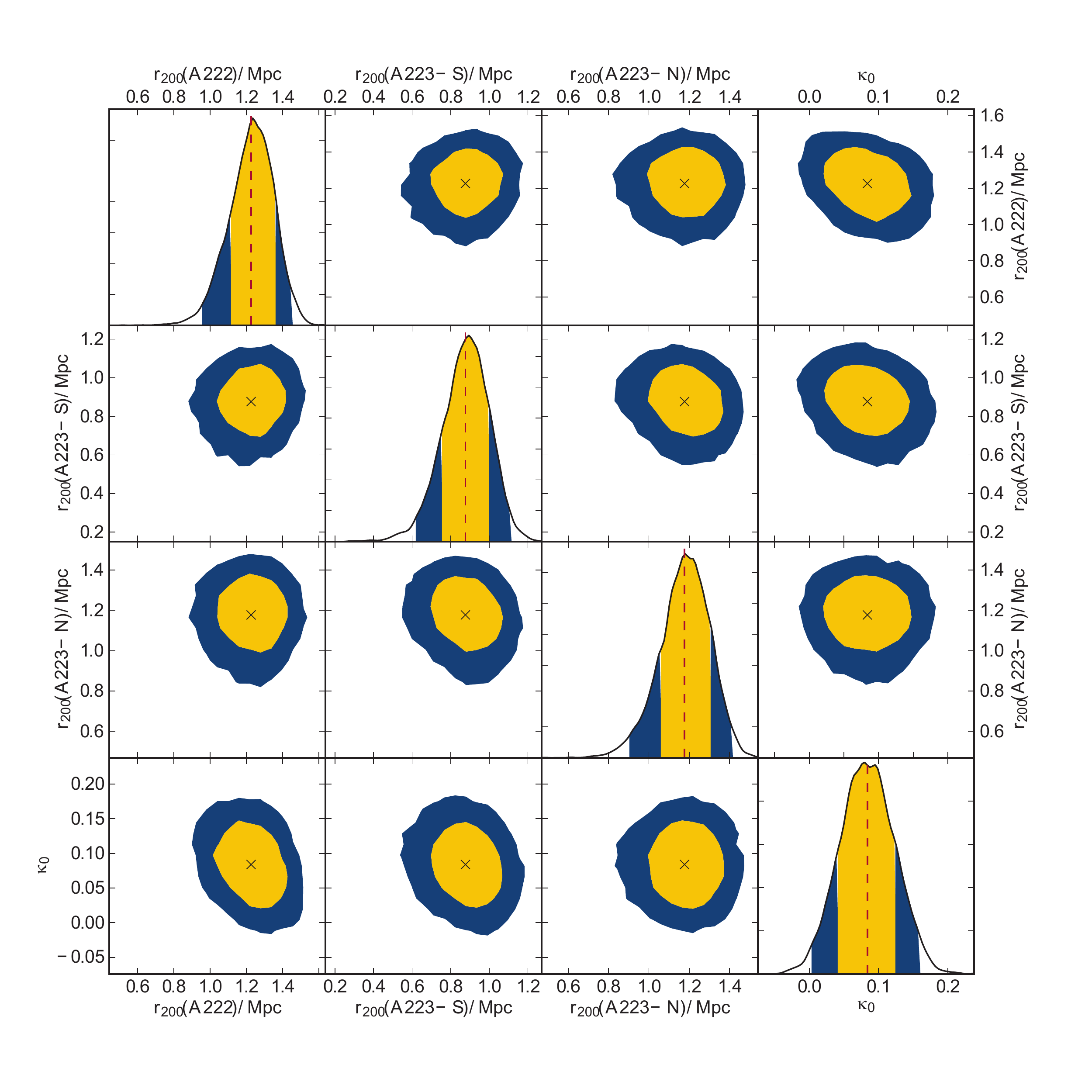}
  \caption{\textbf{Posterior probability distributions for cluster
      virial radii and filament strength.} Shown are the $68\%$ and
    $95\%$ confidence intervals on the cluster virial radii
    $r_{200}(\cdot)$ and the filament strength $\kappa_0$. The
    confidence intervals are derived from 30,000 MCMC sample points.
    The filament model is described by $\kappa(\theta, r) = \kappa_0
    \left\{1 + \exp\left[(|\theta| - \theta_\mathrm{l})/\sigma\right]
      + \left(r/r_\mathrm{c}\right)^2 \right\}^{-1}$, where the
    coordinate $\theta$ runs along the filament ridge line and $r$ is
    orthogonal to it. This model predicts the surface mass density at
    discrete grid points from which we computed our observable, the
    reduced shear, via a convolution in Fourier space. The data cannot
    constrain the steepness of the exponential cut-off at the filament
    endpoints $\sigma$ and the radial core scale $r_\mathrm{c}$. These
    were fixed at their approximate best-fit values of $\sigma =
    0.45$\,Mpc and $r_\mathrm{c} = 0.54$\,Mpc. The data also cannot
    constrain the cluster ellipticity and orientation. These were held
    fixed at the values measured from the isodensity contours of
    early-type galaxies\cite{2002A&A...394..395D}. The ratios of
    minor/major axes and the position angles of the ellipses are
    $(0.63, 0.69, 0.70)$ and $(65\deg, 34\deg, 3\deg)$ for A~222,
    A~223-S, and A~223-N, respectively. We further explore the impact
    of cluster ellipticity on the filament detection in the
    supplementary information.}
  \label{fig:posterior}
\end{figure}

The best fit parameters of this model were determined with a
Monte-Carlo Markov Chain (MCMC) and are shown in
Fig.~\ref{fig:posterior}. The likelihood-ratio test prefers models
with a filament component with $96.0\%$ confidence over a fit with
three NFW halos only. A small degeneracy exists in the model between
the strength of the filament and the virial radii of A~222 and
A~223-S. The fitting procedure tries to keep the total amount of mass
in the supercluster system constant at the level indicated by the
observed reduced shear. Thus, it is not necessarily the case that
sample points with a positive filament contribution indeed have more
mass in the filament area than a 3 clusters only model has. The reason
is that the additional filament mass might be compensated for with
lower cluster masses. We find that the integrated surface mass density
along the filament ridge line exceeds that of the clusters only model
in $98.5\%$ of all sample points. This indicates that the data
strongly prefers models with additional mass between A~222 and A~223-S
and that this preference is stronger than the confidence level derived
from the likelihood-ratio test. The difference is probably due to the
oversimplified model, which is not a good representation of the true
filament shape. The data on the other hand is not able to constrain
more complex models. Extensions to the simple model we tried were
replacing the flat ridge line with a parabola as well as replacing the
King profile with a cored profile leaving the exponent free. The
latter was essentially unconstrained. The parabolic ridge line model
produced a marginally better fit that, however, was statistically
consistent with the flat model. Moreover, the likelihood-ratio test
did not find a preference for the parabolic shape.

The virial masses inferred from the MCMC are lower than those reported
earlier for this system\cite{2005A&A...440..453D}, which were obtained
from fitting a circular two-component NFW model to A~222 and
A~223. Compared to this approach, our more complex model removes mass
from the individual supercluster constituents and redistributes it to
the filament component. Reproducing the two-component fit with free
concentration parameters, which was used in the previous study, we
find $M_{200}(\mathrm{A~222}) = (2.7 ^{+0.8}_{-0.7}) \times
10^{14}\,M_\odot$, which is in good agreement, and
$M_{200}(\mathrm{A~223}) = (3.4 ^{+1.3}_{-1.0}) \times
10^{14}\,M_\odot$, which overlaps the $1\sigma$ error bars of the
earlier study. Here and in the following, all error bars are single
standard deviations.

The detection of a filament with a dimensionless surface mass density
of $\kappa \sim 0.03$ is unexpected. Simulations generally predict the
surface mass density of filaments to be much
lower\cite{2005A&A...440..453D} and not to be detectable
individually\cite{2010MNRAS.401.2257M}. These predictions, however,
are based on the assumption that the longer axis of the filament is
aligned with the plane of the sky and that we look through the
filament along its minor axis. If the filament were inclined with
respect to the line-of-sight and we were to look almost along its
major axis, the projected mass could reach the observed level. A
timing argument\cite{1959ApJ...130..705K,1986ApJ...307....1S} can be
made to show that the latter scenario is more plausible in the A~222/3
system. In this argument we treat A~223 as a single cluster and
neglect the filament component, such that we have to deal only with
two bodies, A~222 and A~223. The redshifts of A~222 and A~223 differ
by $\Delta z = 0.005$, corresponding to a line-of-sight separation of
$18$\,Mpc if the redshift difference is entirely due to Hubble flow.
Let us assume for a moment that the difference is caused only by
peculiar velocities. Then at $z = \infty$, the clusters were at the
same location in the Hubble flow. We let them move away from each
other with some velocity and inclination angle with respect to the
line-of-sight and later turn around and approach each other. The
parameter space of total system mass and inclination angle that
reproduces the observed configuration at $z=0.21$ is completely
degenerate. Nevertheless, in order to explain the observed
configuration purely with peculiar velocity, this model requires a
minimum mass of $(2.61 \pm 0.05)\times10^{15}\,M_\odot$ with an
inclination angle of 46 degrees, where the error on the mass is caused
solely by the uncertainty of the Hubble constant. Since this is more
than 10 standard deviations above our mass estimate for the sum of
both clusters, we infer that at least part of the observed redshift
difference is due to Hubble flow, and that we are looking along the
filament's major axis.

\begin{figure}
  \includegraphics[width=\textwidth]{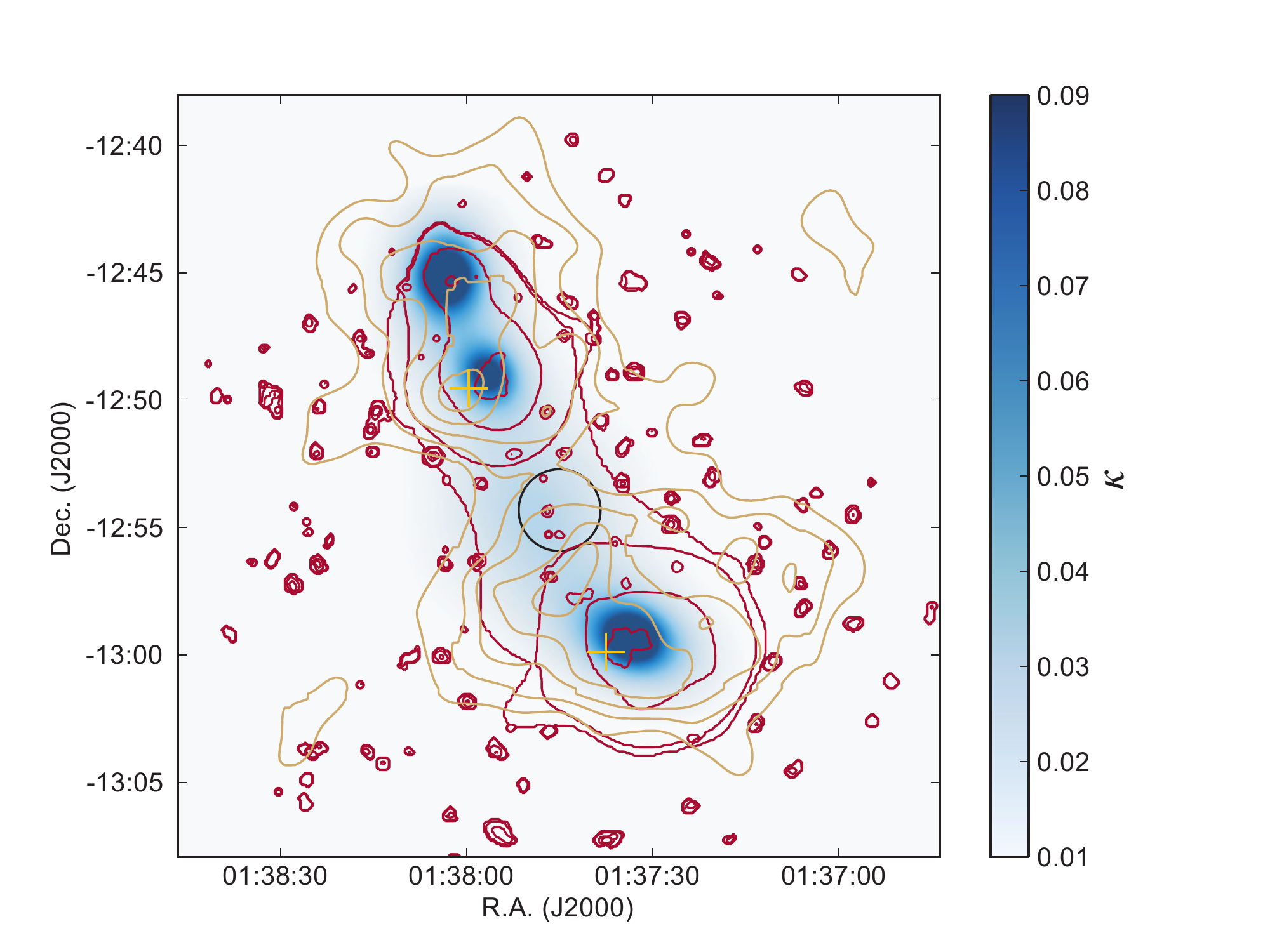}
  \caption{\textbf{Surface mass density of the best fit parametric
      model.} The surface mass density distribution of the best
    parameters in Fig.~\ref{fig:posterior} was smoothed with a
    $2\arcmin$ Gaussian to have the same physical resolution as the
    mass reconstruction in Fig.~\ref{fig:mass-reconstruction}. The
    yellow crosses mark the end points of the filament model. These
    were determined from the visual impression of the filament axis in
    Fig.~\ref{fig:mass-reconstruction}. The MCMC is not able to
    constrain their location. In the model, the filament ridge line is
    not aligned with the axis connecting the centers of A~222 and
    A~223-S. This is a fairly common occurrence ($\sim 9\%$) for
    straight filament but may also indicate some curvature, which
    occurs in $\sim 53\%$ of all intercluster
    filaments\cite{2005MNRAS.359..272C} and is not included in our
    simple model. Overlayed are X-ray contours from XMM-Newton
    observations\cite{2008A&A...482L..29W} (red) and significance
    contours of the colour-selected early-type galaxy
    density\cite{2005A&A...440..453D} (beige), showing the alignment
    of all three filament constituents. The black circle marks the
    region inside which the gas mass and the filament mass were
    estimated.}
  \label{fig:model_fit}
\end{figure}

The combination of our weak-lensing detection with the observed X-ray
emission of $0.91\pm0.25$\,keV WHIM plasma\cite{2008A&A...482L..29W}
lets us constrain the hot gas fraction in the filament. Assuming that
the distribution of the hot plasma is uniform and adopting a
metallicity of $Z = 0.2$ Solar, the mass of the X-ray emitting gas
inside a cylindrical region with radius $330$\,kpc centred on
(01:37:45.00, $−$12:54:19.6, Figure~\ref{fig:model_fit}) with a length
along our line-of-sight of $l = 18$\,Mpc, as suggested by our timing
argument, is $M_\mathrm{gas} = 5.8 \times 10^{12}\,M_\odot$. The
assumption of uniform density is certainly a strong simplification.
Because the X-ray emissivity depends on the average of the squared gas
density, a non-uniform density distribution can lead to strong changes
in the X-ray luminosity. Thus, if the filament consists of denser
clumps embedded into lower density gas (as has been observed in the
outskirts of the Perseus Cluster\cite{2011Sci...331.1576S}), or even
if there is a smooth non-negligible density gradient within the region
used for spectral extraction, then our best fit mean density will be
overestimated. The quoted gas mass should therefore be considered as
an upper limit, and the true mass may be lower by up to a factor of
2--3.

We estimated the total mass of the filament from the reconstructed
surface mass-density map and the model fits within the same region
where we measured the gas mass. The conversion of dimensionless
surface mass density to physical units requires knowledge of the
source redshifts. We randomly sampled galaxies with our
R$_\mathrm{c}$-band magnitude distribution from photometric redshift
catalogues\cite{2006A&A...457..841I}. The mean redshift of these
random catalogues is $z_\mathrm{s} = 1.2$. We emphasize that for a
cluster at $z = 0.21$, the error in mass caused by the uncertainty of
the redshift distribution is small. An error as large as $\Delta
z_\mathrm{s} = 0.2$ causes only a 5\% error. In the reconstructed
$\kappa$-map, the mass inside the extraction circle is $M_\mathrm{fil}
= (6.5 \pm 0.1) \times 10^{13}\,M_\odot$, where the error is small due
to the highly correlated noise of the smoothed shear field inside the
extraction aperture. For the parametric model fit, the inferred mass
is higher but consistent within one standard deviation,
$M_\mathrm{fil} = (9.8 \pm 4.4) \times 10^{13}\,M_\odot$. The
corresponding upper limits on the hot gas fractions vary between
$f_\mathrm{X} = 0.06 - 0.09$, a value that is lower than the gas
fraction in galaxy clusters\cite{2008MNRAS.383..879A}. This is
consistent with the expectation that a significant fraction of the
WHIM in filaments is too cold to emit X-rays detectable by
XMM-Newton\cite{2001ApJ...552..473D}.



\begin{addendum}
 \item [Supplementary Information] is linked to the online version of
   the paper at www.nature.com/nature
\item [Acknowledgements] JPD was supported by NSF grant AST
  0807304. AS acknowledges support from the National Aeronautics and
  Space Administration through the Einstein Postdoctoral Fellowship
  Award Number PF9-00070.
 \item [Author Contributions] JPD led the project, reduced the optical
   data, performed the weak lensing analysis and wrote the manuscript.
   NW contributed to the writing of the manuscript. NW, AF, and AS
   performed the X-ray analysis and estimated the gas mass. The timing
   argument was made by DC. LM and TK wrote the shear estimation
   code. All authors discussed all results and commented on the
   manuscript. 
 \item[Competing Interests] The authors declare that they have no
   competing financial interests.
 \item[Correspondence] Correspondence and requests for materials
   should be addressed to JPD \\(email: jorgd@umich.edu).
\end{addendum}

\newpage
\pagestyle{empty}

\clearpage
\newpage

\clearpage
\newpage

\clearpage
\newpage

\renewcommand{\thefigure}{S\arabic{figure}}
\setcounter{figure}{0}

\renewcommand{\thesection}{S\arabic{section}}
\setcounter{section}{0}

\section*{Supplementary Information}
\label{sec:suppl-inform}

In this document we provide additional information on the data
reduction, shear measurement, and tests performed.

\section{Data}
\label{sec:data}

We reduced the data using the publicly available \textsc{theli}
pipeline\cite{2005AN....326..432E}. The data were astrometrically
calibrated against the USNO-B1.0 catalogue, which has a precision of
$\sim 0\farcs2$. Since this is insufficient to align objects on
dithered exposure without introducing errors in the gravitational
shear measurement, \textsc{theli} uses the \textsc{SWarp} calibration
software to align images internally. The astrometric residuals in this
process are smaller than $0\farcs02$ or a tenth of a pixel.

No photometric standards were observed in the nights our data was
taken. We calibrated our R$_\mathrm{c}$-band data against the catalogue
of our earlier study of this system\cite{2005A&A...440..453D}.  We
fixed the zero-points of the other two bands by matching the expected
colours of stars from the Pickles stellar
library\cite{1998PASP..110..863P} to the observed colours of stars in
the field.

The gravitational shear was estimated independently in all three
passbands using the lensfit
algorithm\cite{2007MNRAS.382..315M,2008MNRAS.390..149K}. Lensfit is a
Bayesian model fitting code, which fits the sum of an exponential and
a de Vaucouleurs profile to galaxies. The model fit has six free
parameters, the galaxy position $(x, y)$, its ellipticity $(e_1,
e_2)$, brightness, bulge-fraction, and galaxy scale $r$. The model
fitting is done in Fourier space so that the brightness and position
can be marginalised over analytically. The radius and bulge-fraction
are marginalised over numerically, leaving a likelihood surface as a
function of $(e_1, e_2)$. Lensfit acts on the individual exposures
and generates a posterior probability for each galaxy by summing over
the posterior ellipticity distributions generated by measurements in
each exposure.

Because lensfit is a forward fitting method, its model space needs to
be convolved with a PSF model describing the stellar point spread
function at the location of each galaxy. We modelled the spatial
variation of the PSF across the focal plane with a bivariate
polynomial, whose zeroth and first order coefficients were allowed to
vary from chip to chip.

Galaxy shapes are highly correlated across different passbands and
simply combining shear estimates from different passbands entails the
risk of degrading them to the band with the worst seeing or the
largest remaining systematics. To maximise the lensing information
from all three passbands we therefore chose a different route. We used
shape measurements from the R$_\mathrm{c}$-band data, which has the
best seeing and is the deepest. For galaxies without
R$_\mathrm{c}$-band shape information, we used i$^\prime$-band
measurements, and in turn V-band measurements were used for galaxies
that also have no i$^\prime$-band shapes.

We used an empirical approach to find an optimal balance between the
rejection of cluster/foreground galaxies and the density of our
lensing catalogue. We placed aperture mass
filters\cite{1996MNRAS.283..837S} on the three known structures in the
field (A~222, A~223-S, A~223-N). For galaxies with two colours
measured, we rejected galaxies inside a parallelogram placed on a
colour-colour diagram. An MCMC algorithm moved the vertexes of this
parallelogram to maximise the aperture mass signal. A similar
procedure was used for galaxies with one colour measured, either $V -
R_\mathrm{c}$ or $R_\mathrm{c} - i^\prime$. A rectangle inside which
galaxies were rejected from the lensing catalogue was placed on a
colour-magnitude diagram. Again, the corners of this rectangle were
moved to maximise the aperture mass signal. For both procedures, a
maximum magnitude, down to which the colour cut was applied, was
determined at the same time. Finally, for galaxies observed in only
one passband, we imposed a magnitude cut in the same way.

This method rejected galaxies with colours consistent with early-type
galaxies at the cluster redshift and showed that fainter galaxies
produce a stronger aperture mass signal than brighter ones. Both
properties are indications that we measure a real cluster signal and
that the redshift scaling of the lensing signal is at least
approximately correct. At the same time we find that colour and
magnitude cuts made to maximise the aperture mass signal of the known
clusters, also lead to an increase in the lensing strength of the
filament. This supports our view that the filament is a real structure
located at the same redshift as the galaxy clusters.

We also note that the filament is seen in mass reconstructions
generated from lensing catalogues of the individual passbands, but at
lower levels consistent with the shallower data. In the
R$_\mathrm{c}$-band reconstruction, a mass bridge is present at the
$3.5\sigma$ level. In the V- and i$^\prime$-band it is seen at the
$2\sigma$ and $2.5\sigma$ level, respectively. We emphasise again that
one cannot simply combine these significances in quadrature because a
large covariance exists between the shear estimates in different
passbands.

\begin{figure}
  \includegraphics[width=\textwidth]{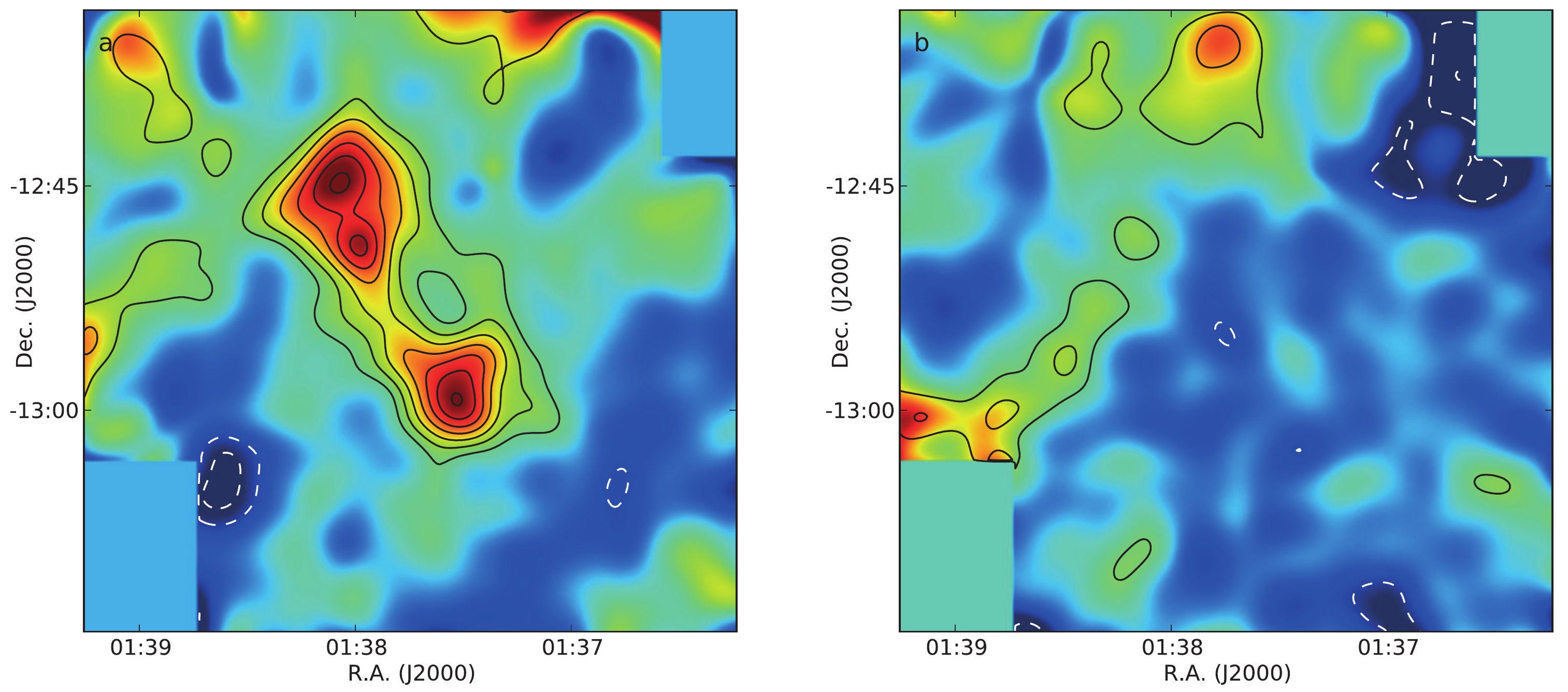}
  \caption{\textbf{E- and B-mode reconstructions.} Panel a again shows
    the mass reconstruction from
    Figure~\ref{fig:mass-reconstruction}. The colour scale in the
    background is dimensionless surface mass density and the contours
    are significance contours starting at $2\sigma$ and rising in
    steps of $1\sigma$ above the mean of the field edges (solid
    black), excluding the top edge, which is strongly affected by
    stray-light. Dashed white contours are at the same negative
    levels. Panel b shows a mass reconstruction of the B-modes,
    obtained by rotating all galaxies by $45\deg$. More and/or higher
    peaks than expected from a Gaussian random field would be an
    indicator of systematic residuals in the shear estimation. The
    peak count here is consistent with expectations for a pure noise
    field\cite{2000MNRAS.313..524V}. The shear field in both panels
    was smoothed with a $2\arcmin$ Gaussian. Both panels have
    identical colour scales.}
  \label{fig:ebmode}
\end{figure}

Finally, we show E- and B-mode reconstructions of our combined lensing
catalogue to test for PSF modelling systematics
(Figure~\ref{fig:ebmode}). 

\section{Fitting elliptical profiles}
\label{sec:fitt-ellipt-prof}

The ellipticity of the clusters connected by a filament has the
potential to be confused with the filament itself and may in extreme
cases even lead to the visual impression of a filament even if the
clusters' mass distribution is completely described by elliptical NFW
halos. In this section we further explore the influence of cluster
ellipticity on the significance of our filament detection.

To recapitulate, we modelled the clusters with fixed ellipticities
measured from the distribution of their early type galaxies. In this
configuration, A~223-S points almost exactly along the filament
ridgeline and A~222 is about halfway turned towards the filament.
Consequentially both clusters contribute more mass towards the
filament area than they would in a spherical configuration. This is
reflected in small differences in the filament significance if we fit
three spherical clusters. In this case, the integrated mass along the
filament exceeds that of the three clusters only case in 98.9\%
compared to 98.5\% for the fixed ellipticities. The likelihood-ratio
test prefers three spherical clusters plus a filament over a model
with only three spherical clusters at 97.8\% confidence, compared to
the 96.0\% significance for the fixed ellipticities used in the main
paper. The filament mass in the case of spherical halos is higher by
about 30\%. This is within the error bars we report but easily
explained by the diminished contribution of A~222 and A~223-S to the
surface mass density in the filament region.

\begin{figure}
  \includegraphics[width=\textwidth]{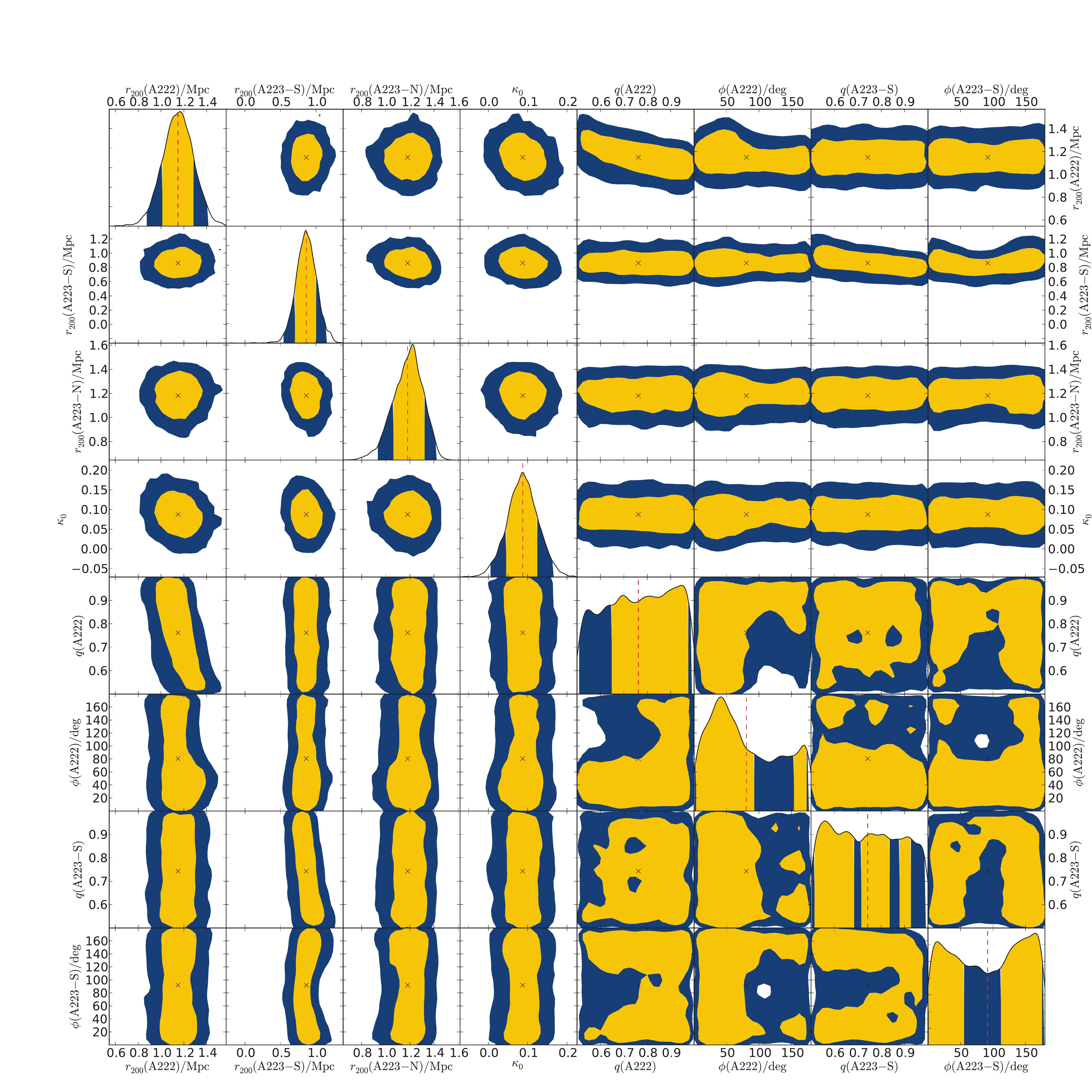}
  \caption{\textbf{Posterior probability distributions with free
      ellipticities.} Similar to Figure~\ref{fig:posterior} we show
    the 68\% and 95\% confidence contours on the cluster virial radii
    $r_{200}(\cdot)$ and filament strength $\kappa_0$. In addition we
    tried to constrain the cluster axis ratios $q(\cdot)$ and position
    angles $\phi(\cdot)$ for A~222 and A~223-S. Their values for
    A~223-N were held fixed at those used in the main paper, as the
    mass distribution of A~223-N shows no degeneracy with the filament
    strength. We put a flat prior $0.5 < q < 1$ on the axis ratio to
    encompass all but the most elliptical
    halos\cite{2005ApJ...629..781K}. This also includes the values
    used in the main paper. Allowing for more extreme axis ratios does
    not seem to be warranted by the absence of a trend in the filament
    strength with axis ratios.}
  \label{fig:posterior_ellipticity}
\end{figure}

We used fixed cluster ellipticities because the data is not able to
constrain the ellipticity, as evidenced by
Figure~\ref{fig:posterior_ellipticity}. However, even though we
allowed for more elongated halo shapes and for A~222 to point exactly
along the filament ridgeline, the marginalised probability
distribution of $\kappa_0$ shows that a positive filament distribution
is still strongly favoured. Again the integrated surface mass density
along the filament exceeds that of a model without filament component
for 98.4\% of all sample points and the likelihood ratio test prefers
the presence of a filament at 98.2\% significance. The latter number
is noticeably higher than what we report for fixed cluster
ellipticities. The reason is that the ``best fit'' values, which are
really unconstrained, have A~223-S oriented perpendicular to the
filament. As a result, the presence of a filament is even more
strongly required than in the configuration we considered in the main
body of the paper. The likelihood-ratio test shows that elliptical
clusters are slightly preferred by the data over spherical ones
(84.3\%), but -- as suggested by the flat posteriors on $q$ -- no
preference exists for the specific values of axis ratios and position
angles chosen by us (19.5\%). In light of what the data tell us, we
thus consider our choice of cluster ellipticities to be conservative.


\newpage

\clearpage
\newpage



\end{document}